\documentclass[aps,prl,reprint,amsmath,amssymb,groupedaddress,showkeys,floatfix,superscriptaddress,showpacs]{revtex4-2}

\usepackage{graphicx}
\usepackage{dcolumn}
\usepackage{bm}
\usepackage{setspace}%
\usepackage[sort&compress]{natbib}%

\usepackage{inputenc}%
\usepackage[integrals]{wasysym}%
\usepackage[normalem]{ulem} \usepackage{color} 
\DeclareUnicodeCharacter{2212}{-}

\newcommand{\BiSe}{Bi$_{2} $Se$_{3}$}
\newcommand{\BiTe}{Bi$_{2} $Te$_{3}$}
\newcommand{\TBS}{TlBiSe$_{2}$}

\newcommand{\Gbar}{\ensuremath{{\bar{\Gamma}}}}
\newcommand{\Kbar}{\ensuremath{{\bar{\text{K}}}}}

\newcommand{\PreserveBackslash}[1]{\let\temp=\\#1\let\\=\temp}

\begin{document}

\title{On the floating of the topological surface state on top of a thick
	lead layer:\\ The case of the Pb/\BiSe\ interface}

\author{Oreste De Luca}
\affiliation{Laboratorio di Spettroscopia Avanzata dei Materiali, STAR IR, Via Tito Flavio, Università  della Calabria, 87036, Rende (CS), Italy}
\affiliation{Dipartimento di Fisica, Universit\`{a} della Calabria,
Via P.Bucci, 87036 Arcavacata di Rende (CS), Italy}

\author{Igor A. Shvets}
\affiliation{Tomsk State University, 634050 Tomsk, Russia}

\author{Sergey V. Eremeev}
\affiliation{Institute of Strength Physics and Materials Science, Russian Academy of Sciences, 634055 Tomsk, Russia}

\author{Ziya S. Aliev}
\affiliation{Baku State University, AZ1148 Baku, Azerbaijan}
 
\author{Marek Kopciuszynski}
\affiliation{Sincrotrone Trieste S.C.p.A., Area Science Park, I-34012 Basovizza, Trieste, Italy}

\author{Alexey Barinov}
\affiliation{Sincrotrone Trieste S.C.p.A., Area Science Park, I-34012 Basovizza, Trieste, Italy}
 
\author{Fabio Ronci}
\affiliation{Istituto di Struttura della Materia-CNR (ISM-CNR), Via del Fosso del Cavaliere, 00133 Roma, Italy}
 
\author{Stefano Colonna}
\affiliation{Istituto di Struttura della Materia-CNR (ISM-CNR), Via del Fosso del Cavaliere, 00133 Roma, Italy}

\author{Evgueni V. Chulkov}
\affiliation{Departamento de Pol\'imeros y Materiales Avanzados: F\'isica, Qu\'imica y Tecnolog\'ia, Facultad de Ciencias Qu\'imicas, Universidad del Pa\'is Vasco UPV/EHU, 20080 San Sebasti\'an/Donostia, Spain}
\affiliation{Donostia International Physics Center (DIPC), 20018 Donostia-San Sebasti\'{a}n, Basque Country, Spain}
\affiliation{Centro de F\'{i}sica de Materiales (CFM-MPC), Centro Mixto CSIC-UPV/EHU,  20018 Donostia-San Sebasti\'{a}n, Basque Country, Spain}
\affiliation{Saint Petersburg State University, 199034 Saint Petersburg, Russia}

\author{Raffaele G. Agostino}
\affiliation{Laboratorio di Spettroscopia Avanzata dei Materiali, STAR IR, Via Tito Flavio, Università  della Calabria, 87036, Rende (CS), Italy}
 \affiliation{Dipartimento di Fisica, Universit\`{a} della Calabria,
Via P.Bucci, 87036 Arcavacata di Rende (CS), Italy}

\author{Marco Papagno}
\affiliation{Laboratorio di Spettroscopia Avanzata dei Materiali, STAR IR, Via Tito Flavio, Università  della Calabria, 87036, Rende (CS), Italy}
 \affiliation{Dipartimento di Fisica, Universit\`{a} della Calabria,
Via P.Bucci, 87036 Arcavacata di Rende (CS), Italy}
		
\author{Roberto Flammini}
\email{Roberto.Flammini@cnr.it}
\affiliation{Istituto di Struttura della Materia-CNR (ISM-CNR), Via del Fosso del Cavaliere, 00133 Roma, Italy}

\date{today}

\begin{abstract}
The puzzling question about the floating of the topological surface state on top of a thick Pb layer, has now possibly been answered. A photoemission study of the interface made by Pb on \BiSe\, for different temperature and adsorbate coverage condition, allowed us to demonstrate that the evidence reported in the literature can be related to the surface diffusion phenomenon exhibited by the Pb atoms, which leaves the substrate partially uncovered. Comprehensive density functional theory calculations show that despite the specific arrangement of the atoms at the interface, the topological surface state cannot float on top of the adlayer but rather tends to move inward within the substrate.
\end{abstract}

\maketitle

\section{Introduction}

Triggering the migration of a topological surface state (TSS) from a substrate into an adsorbate layer is an important goal as it means the topology--protected properties of a substrate could be transferred to a suitable conventional material. Research fields like superconductivity, Majorana physics or spintronics might strongly benefit from the spin-orbit texture or from the resilience to disorder and defects, inherited from the topological matter \cite{Hutasoit_PRB_prox, Hirahara_PRL_2011_Bi_prox, Hirahara_PRL_2012_Bi_prox,Wu2013_prox, Bahramy2012, Yoshimi2014_prox,Essert_2014_prox, Shoman2015_prox, Seixas2015,Soumyanarayanan2016, Costa_ACSomega_prox, Chae_ACSnano_prox, Holtgrewe_SciRep_2020}. 

The subtle details underlying this phenomenon have not been fully explained. Underpinning of the TSS migration is a transformation of the spin--orbit texture of electronic bands of the substrate and of the adlayer due to matching at the interface. Provided that the spin-orbit coupling strength is comparable to other relevant terms of the Hamiltonian describing the topological insulator (TI), \cite{Bahramy2012, Shoman2015_prox, Waugh2016} band hybridization can take place and new physical phenomena may occur, such as the migration of the TSS. \cite{Wu2013_prox, Wang_2013_CaSr, Seixas2015, Soumyanarayanan2016,Holtgrewe_SciRep_2020} 

The first reports on the effects of Pb growth on top of a TI have led to contrasting results. On \BiTe\ D.A.~Estyunin \textit{et al.} \cite{Estyunin2022_Pb} predict the withdrawal of the TSS towards the substrate bulk, while C.X. Trang \textit{et al.} \cite{Trang2020} claim the possibility to transfer the TSS of a \TBS\ substrate up to the 20 monolayer (ML) thick Pb adlayer. If that proved true, there could be a huge step towards the incorporation of topological matter into functional electronic devices \cite{Breunig2022}.

The recent studies about Pb on layered TI reported in the literature mainly focus on the superconducting nature of the interface \cite{Surnin2018, Stolyarov_JPCL_2021} or on the exploitation of its magnetic properties \cite{Estyunin2022_Pb, Klimovskikh_JPCL_2022}. Interestingly, these investigations also shed light on the delicate phase of the growth and on the subsequent formation of a wetting layer (WL), within which the adsorbate and substrate atoms strongly interact forming an alloy of unknown stoichiometry. 

Stimulated by these recent results we undertook an experimental and theoretical study of the Pb/\BiSe\ interface. Scanning photoemission microscopy (SPEM), spatially resolved X-Ray photoelectron spectroscopy (XPS) and angle resolved photoelectron spectroscopy ($\mu$--ARPES) measurements were carried out at the Elettra synchrotron radiation source. Moreover, Density Functional Theory (DFT) calculations were performed considering a sharp interface to predict the localization of the TSS in the case of a thin (1~ML) and of a thick (7~ML) Pb layer to unveil the nature of the interface and of its electronic structure as it starts to form.

\section{Experimental and Computational details}
The \BiSe\ samples are (0001)-oriented single crystals grown by the modified Bridgman method. These samples have been previously characterized by X-ray diffraction. The \BiSe\ surface was prepared by \textit{in-situ} exfoliation, in an ultra-high-vacuum (UHV) chamber. Lead was sublimated from a temperature-controlled Knudsen cell. All depositions were performed at liquid nitrogen (LN) temperature with a deposition rate of 0.33 \AA/min. We considered 1 ML of Pb equal to 0.286 nm, a value corresponding to the interplane distance along the (111) direction.

The measurements were carried out at the Spectromicroscopy beamline \cite{DudinJSR}. The photons were focused through a Schwarzschild objective in order to obtain a sub--$\mu$m beam size spot. For the present measurements, the total energy resolution was set to 70 meV with an angle resolution of 0.6$^{\circ}$. The ARPES maps were acquired at 74 eV and 27 eV photon energies, along the $\Kbar\Gbar\Kbar$ direction. The vertical stripes appearing in the ARPES maps are due to a imperfections of the entrance slit of the detector. The subtraction of such stripes would have led to a loss of relevant information, this is why we decided to display the maps as measured, without any data handling. The stripes do not influence in any way the results and the conclusions of our work. The measurements were performed under UHV in the 10$^{-10}$~mbar range. The SPEM intensity images were obtained by integrating the photoemission spectra within the relative energy window ($\approx$~3.5~eV). The XPS spectra were acquired at 74 eV and angle integrated, with an acceptance angle of $\pm$~7.5$^{\circ} \times \pm0.3^{\circ} $ corresponding to $\pm$ 0.3 \AA$^{-1} \pm$ 0.012 \AA$^{-1} $, around the $\Gbar$ point. The photoemission measurements were performed at LN temperature. 

Electronic structure calculations were carried out within the density functional theory using the projector augmented-wave (PAW) method \cite{Blochl.prb1994} as implemented in the VASP code \cite{vasp1,vasp2}. The exchange-correlation energy was treated using the Perdew-Burke-Ernzerhof (PBE) type generalized gradient approximation (GGA) \cite{Perdew.prl1996}. The Hamiltonian contained scalar relativistic corrections and the spin-orbit coupling was taken into account. In order to describe the van der Waals interactions we made use of the DFT-D3 approach \cite{Grimme2011}. Spin-orbit coupling was always included when performing relaxations. For analysis of interactions between Pb atoms and the Bi$_2$Se$_3$ substrate we used the projected crystal orbital Hamilton population (pCOHP) method \cite{Dronskowski1993,Deringer2011} implemented within Local-orbital basis suite towards electronic structure reconstruction code \cite{LOBSTER-2016}.
To simulate the Pb films on the Bi$_2$Se$_3$-$1\times 1$ surface we used a slab consisting of 8 quintuple layers (QLs) of the substrate with the PBE-optimized bulk lattice constants with Pb layers placed on both slab surfaces. The atomic positions of Pb as well as  atoms of the topmost QL of Bi$_2$Se$_3$ were optimized. For consideration of the $6/5 \times 6/5$ Pb(111)/Bi$_2$Se$_3$ models the supercells composed of the single Pb(111)-$6\times 6$ layer on top of the single QL of Bi$_2$Se$_3$-$5 \times 5$ was used. The geometry optimization was performed until the residual forces on atoms became smaller than 1 meV/\AA. The 10$\times10\times$1 and 2$\times2\times$1 $k$-point meshes were used to sample the $1\times 1$ and $5 \times 5$ surface Brillouin zones, respectively.

\section{Results and Discussion}

\begin{figure}[t]
		\includegraphics[width=\columnwidth]{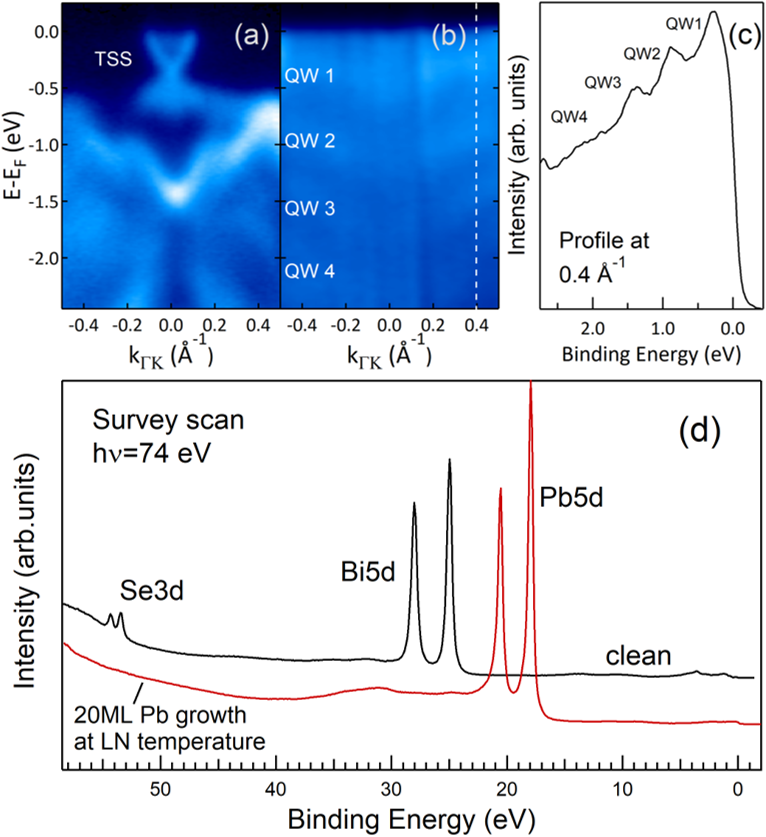}
	\caption{(a) ARPES map of the pristine \BiSe\ surface along the \Kbar\Gbar\Kbar\ high symmetry direction. At \Gbar\ the Dirac point and the TSS are noted. (b) ARPES map of the same sample surface after deposition of 20~ML of Pb. QW states are visible without any feature related to the TSS. (c) Line profile (EDC) corresponding to the dashed line of panel (b). In (d) the corresponding XPS survey scans are shown. All spectra have been taken at h$\nu$=74 eV. The zero binding energy corresponds to the Fermi level.} \label{Fig1}
\end{figure}

In Fig.~\ref{Fig1}(a) the ARPES map displaying the band structure of the pristine substrate is shown. The TSS and the Dirac point (DP) at about --0.3 eV below the Fermi level are readily visible at \Gbar, along with the bulk valence bands \cite{Zhang2009,Kuroda_PRL_2010}. Fig.~\ref{Fig1}(b) features the ARPES map recorded after deposition of 20 ML of Pb. 
The several electron--like parabolas with vertex at \Gbar\ are attributed to, at least, four  quantum well states (QWS) \cite{WeiPRB_2005,MansPRB_2005}, demonstrating the flatness and uniformity of the adlayer, owing to the growth protocol. In Fig.~\ref{Fig1}(c), the energy distribution curve (EDC) extracted from Fig.~\ref{Fig1}(b) better highlights the features corresponding to the QWS. The complete burial of the substrate surface is demonstrated by the corresponding survey scan displayed in Fig.~\ref{Fig1}(d). After the Pb growth no traces of Bi or Se were recorded. The most noticeable feature, however, is the absence of the TSS in Fig.~\ref{Fig1}(b). 

\begin{figure*}[t]
		\includegraphics[width=\textwidth]{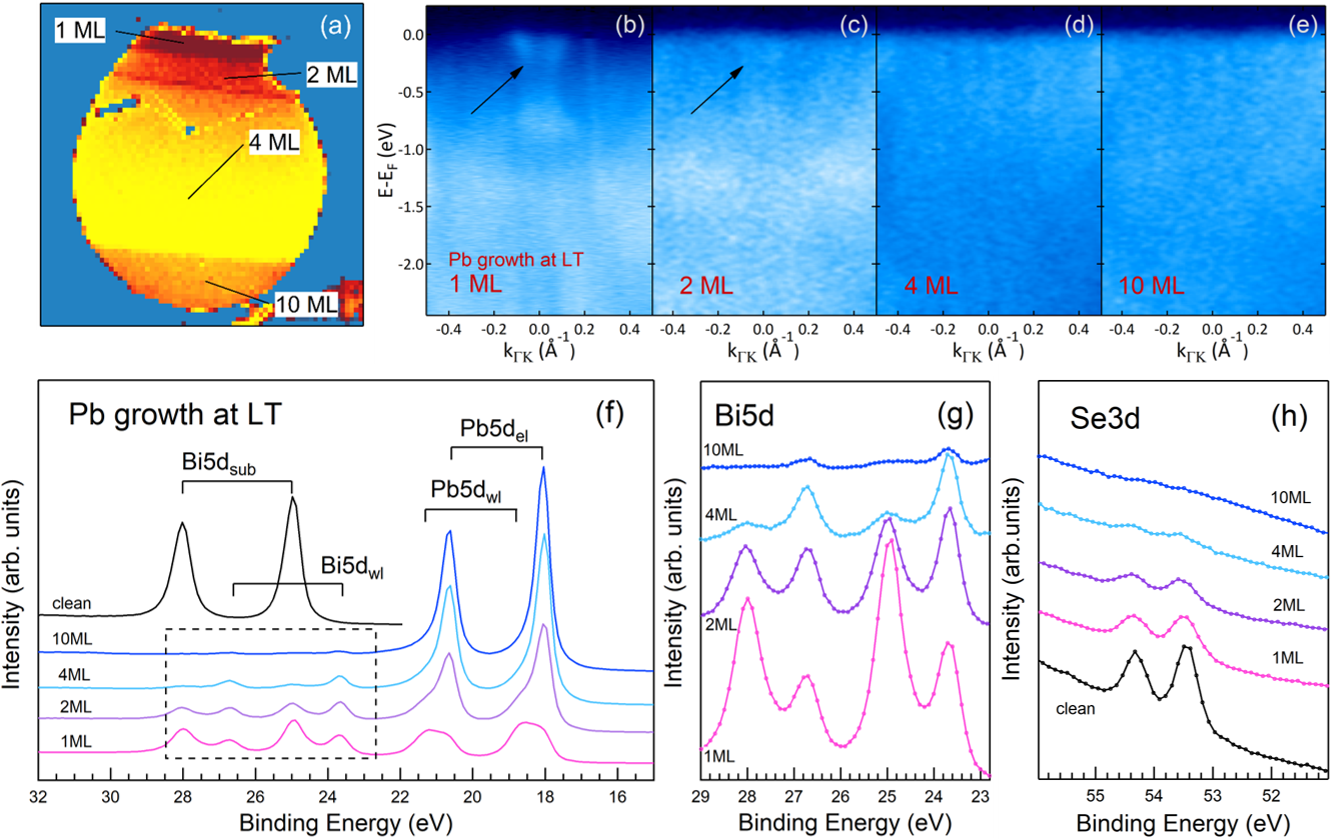}
	\caption{(a) SPEM intensity map ($6000 \times 6000$ $\mu$m$^2$) taken on the Pb-5$d_{5/2}$ core level at about 18 eV binding energy. The areas exhibiting different adsorbate thickness are displayed in false colors. The ARPES images (b-e) and the core level spectra (f-h) were taken on the corresponding zones of the Pb wedge, at $h\nu = 74$ eV photon energy. The dashed rectangle highlights the blow-up shown in panel (g). Panel (h) features the corresponding Se-$3d$ spectra.}
	\label{Fig2}
\end{figure*}

This result suggests that either the topological protection is destroyed or the TSS is buried under the thick Pb adlayer. To this aim, a Pb wedge was fabricated on the \BiSe\ substrate, as the TSS migration process may have taken place below the probing depth of the photoemission process. In Fig.~\ref{Fig2}(a) a SPEM intensity image of the wedge is displayed. The wedge is composed of four areas with 1, 2, 4 and 10 ML of Pb, where $\mu$-ARPES spectra were collected. The use of the SPEM allowed us to pick up the different areas of the wedge without spatial superposition. 
A notable difference is seen between Fig.~\ref{Fig2}(b,c) where traces of the \BiSe\ TSS are recorded, and Fig.~\ref{Fig2}(d,e), where the TSS is, at variance, absent. The QWS are not detected, as they start to form for larger deposits \cite{Upton2004, Upton2005}. 

The core levels in Fig.~\ref{Fig2}(f) feature the Pb-5$d$ doublet consisting of two components of which one, at lower binding energy, can be attributed to the Pb atoms in their bulk environment (Pb$_{el}$, where `\textit{el}' stands for `\textit{elemental}') and the other (Pb$_{wl}$), at higher binding energy, attributed to the WL, where a strong interaction of the Pb atoms with the substrate is expected \cite{Klimovskikh_JPCL_2022,Estyunin2022_Pb}. The Bi-5$d$ doublet shows two components as well: one is attributed to the \BiSe\ environment, Bi$_{sub}$, while the other, at low binding energy, to the wetting layer (Bi$_{wl}$) \cite{Li_2015_alloyPbBi, Klimovskikh_JPCL_2022, Estyunin2022_Pb}. The overall intensity of the Bi-5$d$ peaks decreases and becomes negligible as the amount of Pb increases, demonstrating that an effective burial of the substrate occurs only for high coverage. It is known from the literature that Pb follows the ``electronic growth" mode: flat-topped islands start to form whose height and stability are driven by the compromise between the quantum size effect and adsorbate kinetics.  \cite{MenzelPRB_2003,BuddePRB2000_QSE,Jia2006_electPb}

In the blow-up of the dashed square area displayed in Fig.~\ref{Fig2}(g), it is noticed that the intensity of the WL component increases comparing it to the substrate component and eventually becomes dominant. Our hypothesis is that either residual areas of the WL remain uncovered or a segregation process of Bi atoms on top of the Pb layer occurs. 
The Se-3$d$ core levels shown in Fig.~\ref{Fig2}(h) on the other hand, confirm that the \BiSe\ is fully covered by the Pb layer at higher coverage.		

\begin{figure*}[t]
		\includegraphics[width=0.8\textwidth]{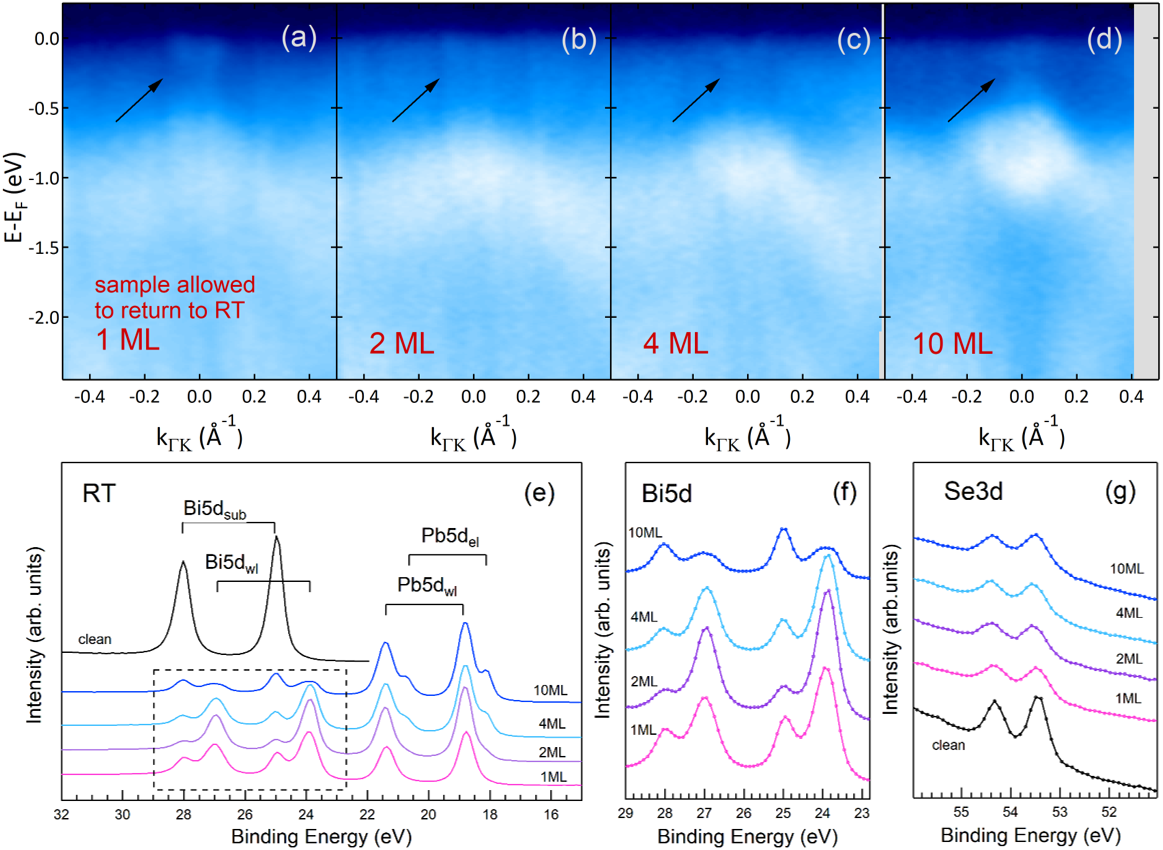}
	\caption{In panels (a-d) the ARPES maps and in (e-g) the core level spectra taken on the Pb wedge after leaving the sample to return to RT. h$\nu$=74~eV photon energy. The arrows indicate the faint TSS of the substrate. The dashed rectangle highlights the blow-up shown in panel (f). In panel (g), the corresponding Se-$3d$ core level spectra are displayed.}
	\label{Fig3}
\end{figure*}

Yet, as is well known, when a Pb layer is prepared at low temperature and then brought to room temperature (RT), a diffusion process occurs, inducing a reduction of the surface coverage \cite{Jia_book_Springer2011}. In order to verify this phenomenon, the sample was allowed to return to RT, keeping it on a carousel under UHV. After 12 hours the sample was checked for contamination, and new measurements were performed. In Fig.~\ref{Fig3}(a-d), the ARPES maps taken on the same wedge areas as before are displayed. All areas show some intensity in correspondence to the \BiSe\ TSS, as indicated by the black arrows.

In the corresponding core level spectra displayed in Fig.~\ref{Fig3}(e), it is noticed that Pb$ _{el}$ almost disappears for low coverage while surviving for a coverage higher than 4 ML; The intensity of the Pb$ _{wl}$ component is clearly larger than Pb$ _{el}$ up to 10 ML. Comparing the Bi-5$d$ doublet components in Fig.~\ref{Fig3}(f) with those of Fig.~\ref{Fig2}(g), we notice a reversed evolution of Bi$_{sub}$ and Bi$_{wl}$. For low coverage, the Bi$_{sub}$ of Fig.~\ref{Fig3}(f), is completely transformed in Bi$_{wl}$, owing to the increased Bi-Pb chemical reaction. Increasing the coverage, the Bi$_{wl}$ component loses intensity in favour of Bi$_{sub}$. This can only be explained by the formation of large uncovered areas. In the literature, this phenomenon was reported in the case of thin films of Pb grown on both insulators and metals (namely Pb/Si and Pb/Cu), where the increase of the island height occurs at the expense of the surface coverage, as verified by photoemission \cite{Upton2004,Dil_2004_Pb}, spatially resolved techniques \cite{Hupalo_2001_Pb,Calleja_Pb_2009}, and explained by a model based on the confinement of free electrons to a quantum well. \cite{Czoschke_2004_Pb_Th}. 

Accordingly, the Se-3$d$ doublet in Fig.~\ref{Fig3}(g), does not show any change in the lineshape or intensity, suggesting a poor interaction with the lead atoms. This can be rationalised by considering a chemically shifted component below our energy resolution or showing a negligible intensity. In the case of the Pb/MnSb$_{2}$Te$_{4}$ \cite{Klimovskikh_JPCL_2022} and at the Pb/Mn(Bi$_{0.7}$Sb$_{0.3}$)$_{2}$Te$_{4}$ \cite{Estyunin2022_Pb} interfaces, the PbTe bond does form, retaining the --2 valence state, and thus not showing any chemical shift with respect to the Te in the \BiTe\ substrate \cite{Klimovskikh_JPCL_2022}. Moreover, the photoionization cross-section of the Se-3$d$ with respect to the Bi-5$d$ results 10 times smaller \cite{YEH19851}.

The evolution of the valence bands and of the loss spectra, as shown in Fig.~S1 and Fig.~S2 of the  Supplemental Material~\cite{[{See Supplemental Material at }]supp}, seems to confirm that the surface gets uncovered: during the growth at low temperature the spectra gradually show the features of an elemental Pb layer; Instead if the sample is left to return to RT, the spectra show evidence again of the Bi and Se outer shells \cite{Thuler1982}. We also notice the photoemission intensity just below the Fermi level that, starting from the peak related to the TSS of the substrate, evolves in a strong metallic behaviour upon growth at low temperature, eventually decreasing at RT (Fig.~S1 \cite{supp}, black arrows). 

Some metals, such as Ag or Au, when deposited using the two--step procedure (where firstly the deposition is made at low temperature and subsequently the sample is allowed to return to RT), undergo an improvement of their flatness and uniformity \cite{CHIANG2000181}. Pb is an exception \cite{ZHANG2005L331,Upton2005} and the above results represent another hint that its peculiar behaviour at different temperatures can explain the detection of the TSS at RT after growth of a thick Pb deposit. In this regard, we cannot exclude the hypothesis of a collective motion of the lead atoms in contrast to the classical nucleation: an explosive island formation could take place owing to the large mismatch (18\%) between the Pb(111) and the \BiSe(0001) lattices \cite{ManPRL2013,HershbergerPRL2014,JAROCH2019125137}. However, the presence of the WL could mitigate the lattice strain and should be taken into account.

	\begin{figure*}[t]
		\includegraphics[width=0.75\textwidth]{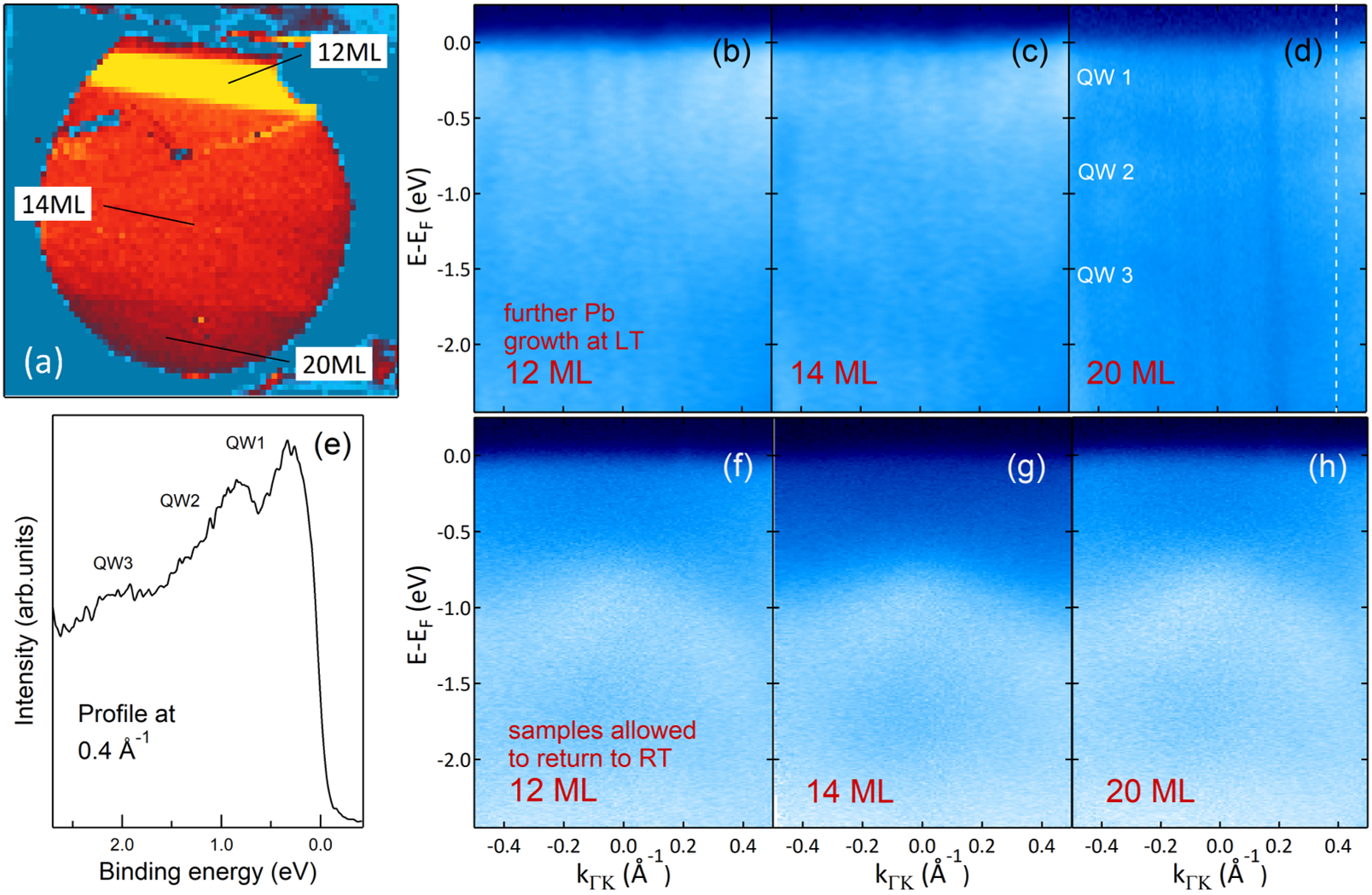}
		\caption{(a) SPEM intensity map of the wedge after an additional deposit of 10~ML at LN temperature. The image, displayed in false colors, was measured on the Pb-5$d_{5/2}$ core level peak and acquired at $h\nu = 74$ eV photon energy. The size of the image is of $6000\times6000$ $\mu$m$^2$. Panels (b-d) and (f-h), ARPES maps recorded at low and room temperature, respectively. In panel (e) evidence of 3 QWs as from the EDC taken at 0.4 \AA$^{-1}$ in panel (d).} \label{Fig4}
	\end{figure*}

In order to ascertain the influence of the temperature, we deposited further 10 ML at LN temperature, trying to bury again the whole interface. In Fig.~\ref{Fig4}(a) the image of the novel wedge is displayed. It is now formed by only three zones as the thin one on top was too small to allow a reliable measurement. The ARPES results displayed in Fig.~\ref{Fig4}(b-d) show no evidence of the TSS. We emphasize that at 74 eV photon energy, the technique is extremely sensitive to the surface, probing only few \AA\ of the Pb/Bi2Se3 interface. The 20 ML region, though, displays QWS, as proved by the EDC presented in Fig.~\ref{Fig4}(e). As a matter of fact, if the deposit as well as the measurements, are carried out with no increase of the temperature, the burial of the TSS is again ensured. By leaving the sample to return to RT, the maps of Fig.~\ref{Fig4}(f--h) show only blurred features recalling the substrate bulk valence band of the pristine surface (below --0.5 eV) and still no TSS.
	
\begin{figure*}[t]
	\includegraphics[width=.7\textwidth]{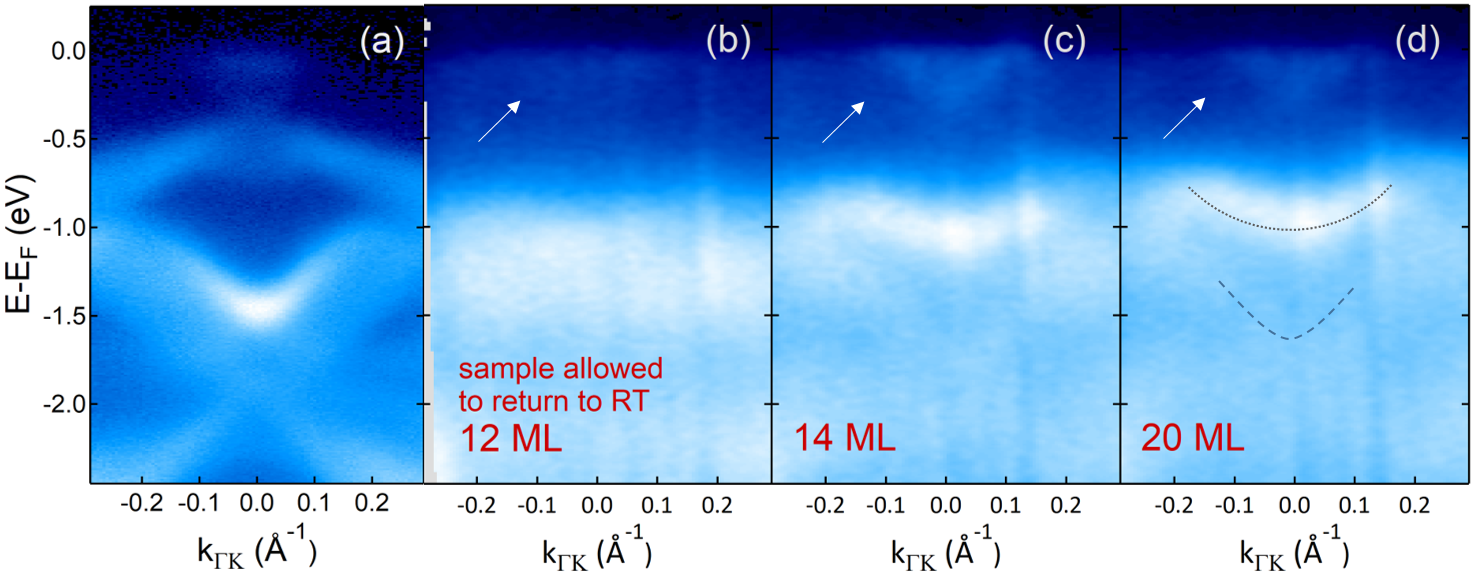}
	\caption{ARPES maps taken at 27 eV photon energy on the pristine sample (a) and on the three different zones relative to the wedge of Fig.~\ref{Fig4}(a), corresponding to 12~ML, 14~ML and 20~ML, panels (b,c,d), respectively. The sample was allowed to return to RT. See text.}
	\label{Fig5}
\end{figure*}

With the aim to understand what happens to the interface we used a photon energy of 27 eV so as to increase the bulk sensitivity. The result is shown in Fig.~\ref{Fig5}~: panel (a) features the map recorded on the pristine surface for comparison, while in panels (b--d) a clear evidence of the TSS is shown again, demonstrating incontrovertibly that the TSS is not floating on top of the Pb layer, but survives underneath. In Fig.~\ref{Fig5}(d), we also attribute the ``V-shape” feature (grey dashed line) to the \BiSe\ bulk valence band. At lower binding energy, we notice a bright feature displaying an electron-like curvature (dotted line), absent in the pristine surface. We tentatively attribute this band to a QW state. Paradoxically, the TSSs in the present case are more evident with respect to the low coverage case of Fig.~\ref{Fig3}(a-d), suggesting an increase of the uncovered areas. The reason can be related to the elastic energy accumulated by a thicker adlayer that overcomes the interaction of the Pb atoms with the substrate, favouring the dewetting phenomenon. One can argue that QSE effects are better seen for smaller height islands and for lower temperatures due to the increased energy stability however, in the present case, due to the strong interaction with the substrate, a larger amount of Pb is needed to form pure and crystalline islands, as part of the adsorbate atoms compose the WL.

It is remarkable that the maps of Fig.~\ref{Fig5}(c) and (d) reproduce well enough the map reported by C.X. Trang \textit{et al.} \cite{Trang2020} for 17 ML of Pb on TlBiSe$_{2}$, where the TSS survives among the QWS. Our interpretation of the data based on the probing depth of the photoelectrons however, does not invoke band hybridization, but rather the dewetting phenomenon. The electronic structure most probably results from the overlap of the features relative to the flat lead islands with those relative to the substrate. Nonetheless, a question remains on the precise localization of the TSS.

\begin{figure*}
	  \includegraphics[width=\textwidth]{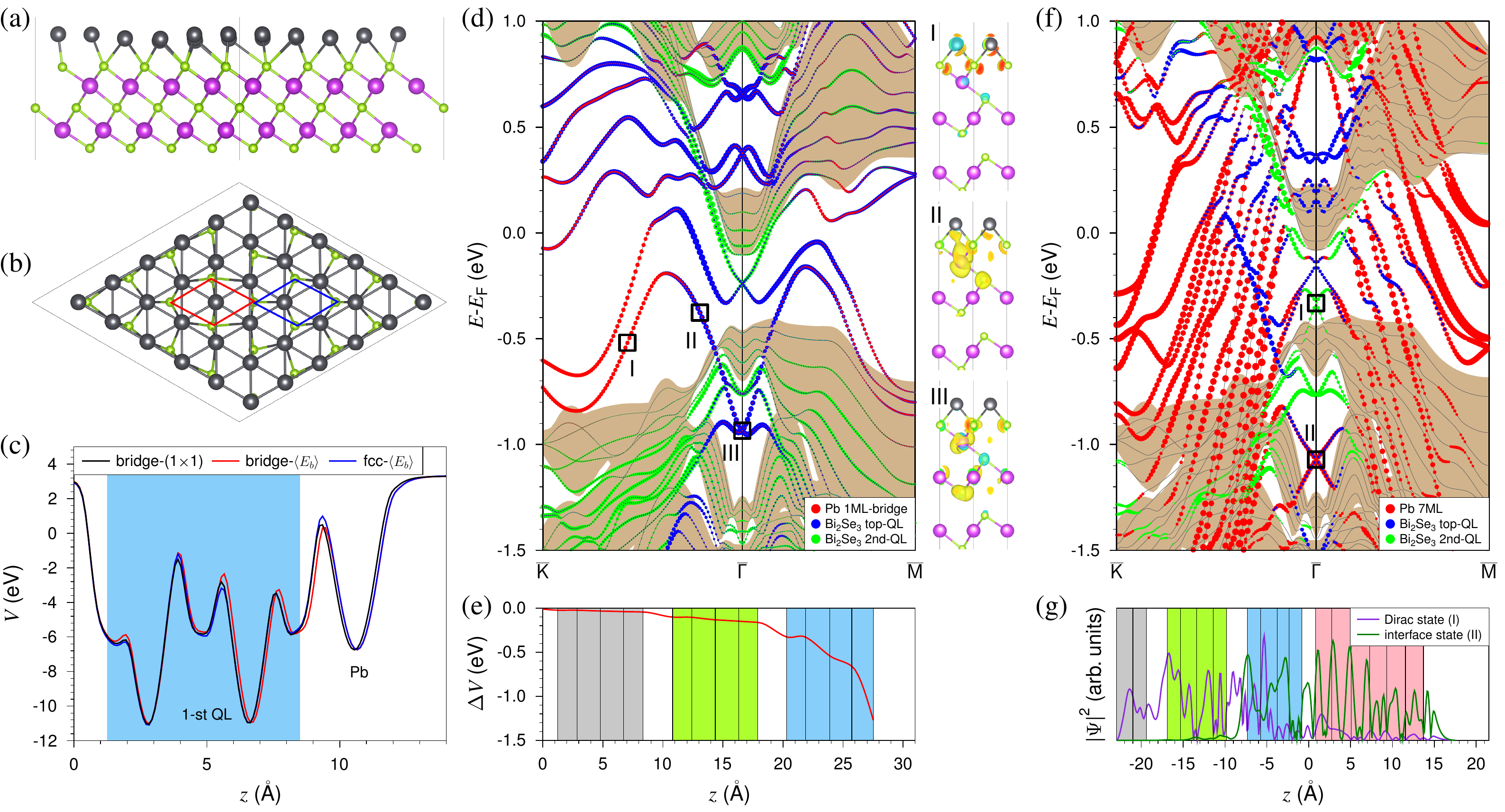}
	\caption{(a) Side view of the relaxed 1 ML Pb/Bi$_2$Se$_3$-$6/5\times 6/5$ interface (one of three models is shown, see text). (b) Top view of the structure; Only two topmost (Pb and Se) layers are shown. (c) In-plane averaged electrostatic potentials $V_z$ for Pb/Bi$_2$Se$_3$ with Pb in bridge position obtained within $1\times 1$ relaxed cell and for those with interatomic distances extracted from $6/5\times 6/5$ model with Pb in bridge (br-$\langle E_b \rangle$) and fcc-hollow (fcc-$\langle E_b \rangle$) positions  (red and blue rhombuses in (b), respectively). Zero $z$ corresponds to the middle of the first vdW gap. (d) Surface electronic structure of the 1 ML Pb/Bi$_2$Se$_3$ interface with Pb in the bridge position. Spatial distributions of the surface states I, II, and III marked with  black squares are shown in right outsets. (e) Potential bending $\Delta V_z$ within the Bi$_2$Se$_3$ slab ($z=0$ corresponds to the middle of the third vdW gap). (f) The electronic spectrum of 7 ML Pb/Bi$_2$Se$_3$. Black squares mark the Dirac-like TSS (I) and the interface state (II) for which the in-plane averaged charge densities are shown in panel (g); $z=0$ corresponds to the interface plane. }
	\label{fig_DFT}
\end{figure*}

To clarify the interaction of the adsorbed layer with the substrate and reveal how the formation of the Pb/Bi$_2$Se$_3$ interface affects the TSS we performed DFT calculations. Considering 1 ML of Pb(111) placed on top of the Bi$_2$Se$_3$ surface within a $1 \times 1$ cell we found that the most favored geometry is that when Pb atom is in the fcc-hollow position. The next preferred geometry is the bridge position having total energy higher by 120 meV/cell. The hcp-hollow and most unfavorable top positions follow. However, the Pb ML stability determined in this way cannot be relied upon, since the lateral lattice constant of Pb(111) is by $\approx$18~\% smaller than that of Bi$_2$Se$_3$ and hence the Pb ML placed on the substrate within a $1\times 1$ cell is unrealistically expanded. 

To overcome the huge lattice mismatch we constructed the ($6 \times 6$)/($5 \times 5$) (briefly $6/5 \times 6/5$) superlattice models simulating incommensurate Pb/Bi$_2$Se$_3$ interface (Pb(111) - $6\times 6$ on top of the Bi$_2$Se$_3$ -- $5 \times 5$) for which the initial Pb--Pb interatomic distance (3.44 \AA) is compressed only by $\approx 1.7$~\%, with respect to the experimental one (3.5 \AA). Three possible ($6/5 \times 6/5$) models, corresponding to A~(0,0), B~($\frac{1}{3}$,~$\frac{2}{3}$), and C~($\frac{2}{3}$, $\frac{1}{3}$) hexagonal layers of Pb(111), were constructed. The relaxed structure of the latter model is shown in Fig.~\ref{fig_DFT}(a,b). After relaxation, the Pb--Pb distances vary from 3.37 to 3.64 \AA\ and the Pb layer acquires a corrugation with a rippling of $\Delta_z\approx 0.75$~\AA\ (see side view in the Fig.~\ref{fig_DFT}(a)). In these superlattice models the Pb atoms appear in more or less distorted "top" (with single Pb-Se bond), "bridge" (where Pb atoms are two-fold coordinated with surface Se atoms) and "hollow" (three-fold coordinated) positions. Estimation of Pb--Se bonding energies via integrated projected crystal orbital Hamilton population (IpCOHP) calculations \cite{Dronskowski1993,Deringer2011,LOBSTER-2016} shows that the bond strength between Pb and neighboring Se atoms can vary in a wide range: from 1.77 (1.97, 2.10) eV to 3.14 (3.56, 3.65) eV for A(B,C)-derived ($6/5 \times 6/5$) models. Remarkably, there is no correspondence between the number of nearest Se neighbors and the value of the bonding energy. For example, three-fold coordinated Pb atoms can be found both at the lower and upper limits of this range. The mean Pb--Se bond energy $\langle E_b \rangle$ averaged over 36 Pb atoms of each ($6/5 \times 6/5$) model and over three models, amounts to 2.55 eV. The Pb atoms possessing the Pb--Se bond energies closest to $\langle E_b \rangle$ are found in distorted top and bridge positions. Among the three-fold coordinated Pb atoms also can be found those having bond energy relatively close to $\langle E_b \rangle$. Two of such Pb atoms, in "bridge" and "fcc-hollow" positions are outlined in Fig.~\ref{fig_DFT}(b) with red and blue rhombuses, respectively. 

Extracting the local interlayer distances within these $1\times1$ rhombuses (averaging the $z$ coordinates in the Se and Bi layers, which also have a small rippling), we constructed the $1\times 1$ cells with Pb in bridge and fcc-hollow positions (br-$\langle E_b \rangle$ and fcc-$\langle E_b \rangle$, respectively). The in-plane averaged electrostatic potentials $V_z$ for these artificial slabs are shown in Fig.~\ref{fig_DFT}(c). As can be seen, they are close to each other as well as close to the $V_z$ of the cell with the Pb atom in the bridge position. In the latter case, the positions of Pb layer and of the topmost Bi$_2$Se$_3$ QL were obtained directly within relaxation of the $1\times1$ cell (br-($1\times1$)). In contrast, the comparison of fcc-$\langle E_b \rangle$ and fcc-($1\times1$) electrostatic potentials highlights the large difference when considering the topmost QL (Fig.~S3(a) \cite{supp}). This is due to the difference between the Pb--Se interlayer spacing obtained from ($6/5 \times 6/5$) model (2.29~\AA) and from the relaxation of the overstretched Pb(111) layer (1.86 \AA). Thus, one can conclude that although the bridge position (within $1\times 1$ cell) is energetically the second preferred one, it represents much better the average bonding at the incommensurate interface between 1 ML Pb(111) and the Bi$_2$Se$_3$ surface. 

The electronic band structure calculated for the 1 ML Pb/\BiSe\ interface with the Pb atom in the bridge ($1\times1$) position, is shown in Fig.~\ref{fig_DFT}(d). Apart from the bulk projected bands shaded in tan, the weights of the surface states are shown by circles of different colours corresponding to the localization in the Pb layer, in the first and in the second QLs (as shown in the key). In order to visualize the position of the maximum density of the states within the atomic lattice we added three outsets, corresponding to the three black boxes in the band structure (labelled I, II and III). 

As can be seen in Fig.~\ref{fig_DFT}(e), due to the large potential bending $\Delta V_z$, obtained by subtracting the potentials of Bi$_2$Se$_3$ with and without the Pb adlayer, the states localized in the first, and even in the second QL, result to be shifted downward in energy. As a consequence, the DP of the former TSS inherited from the Bi$_2$Se$_3$ substrate falls in the local gap, at about $-1$~eV binding energy. 
This state is spread within the topmost QL in the close vicinity of the DP, then it mixes with the Pb states far from the Brillouin zone center, where it experiences a strong localization on the Se and Pb atomic layers (see outsets in Fig.~\ref{fig_DFT}(d)).

It is remarkable that the electronic spectrum of the less realistic fcc-($1\times 1$) structure, leads to very similar results (Fig.~S3(b) \cite{supp}). Nonetheless, in this case, the Pb and Bi$_2$Se$_3$ states are mixed to a greater extent. Thus, it can be concluded that regardless of the position of the Pb monolayer on the Bi$_2$Se$_3$ surface (and one can expect the same effect in the incommensurate Pb layer), the potential bending shifts the former TSS into a local gap, in striking contrast with the ARPES observation, where the TSS remains in the principal gap. The latter should indicate that the continuous Pb monolayer is not formed, but a more complex interaction of Pb with the TI surface takes place.

Next, we considered a relatively thick Pb film of 7 ML (Fig.~\ref{fig_DFT}(f)). The spectrum contains a large number of Pb-film states, which cannot be compared with the dispersion of the measured Pb quantum well states, since the Pb adlayer is overstretched in this geometry and we will focus here on the states of the Bi$_2$Se$_3$ substrate. 
Similarly to earlier studies on Bi$_2$Se$_3$ and other related TIs \cite{Eremeev2013, Wu2013_prox, Menshov2014, Eremeev2015, Menshov2015, Menshchikova2020}, the formation of the interface results in the appearance of an interface state along with the development of a new Dirac state in the principal band gap which departs from the interface region and moves back to the second QL. Indeed, we see the interface state showing the Dirac-like crossing dispersion in the local gap at about $-1$~eV. Traces of this state can be followed up the Fermi level for large $k_\|$, where it overlaps with the set of Pb states and undergoes a localization between the topmost QL and the adlayer. 

The new TSS with the DP is observed in the middle of the principal band gap (at about $-0.35$~eV), mostly localized in the second QL, but decaying outwards and penetrating even into the Pb film (Fig.~\ref{fig_DFT}(g)). The precursor of this state is visible below the bulk conduction band at $\approx -0.2$ eV, in the spectrum of the heterostructure with a single Pb layer (Fig.~\ref{fig_DFT}(d)). Therefore, we can conclude that the TSS of the Bi$_2$Se$_3$ substrate can survive below the relatively thick Pb adlayer, by shifting into the second QL regardless of the peculiarities of the atomic structure at the interface.

All results reported above lead to a fundamental question on whether a TI made by V-VI layered compounds coupled with Pb, can represent a suitable system to induce the migration of the TSS. Apart from the issues related to the stability of the a nanometric-thin Pb layer at different temperatures, the formation of a WL possibly made of a disordered ternary alloy can strongly affect the electronic structure. Ultimately, even in the case of a sharp model interface, the main condition for the relocation of the TSS in the adlayer, is the absence of an interface potential.

\section{Conclusions}

A joint theoretical and experimental study of the behaviour of the TSS at the Pb/\BiSe\ interface was carried out. DFT calculations reveal that the TSS of the substrate escapes the interface region moving downward into the bulk. The detection of the TSS after lead deposition should be attributed to the uncovering of the TI substrate, as a result of surface diffusion of the lead atoms, as demonstrated by the core-level and angle resolved photoemission measurements done for different temperature and adsorbate coverage conditions.

\section{Acknowledgements}
S.V.E. acknowledges support from the Government research assignment for ISPMS SB RAS, project FWRW-2022-0001. I.A.Sh. gratefully acknowledges ﬁnancial support from the Ministry of Education and Science of the Russian Federation within State Task No. FSWM-2020-0033. E.V.C. acknowledges support from Saint Petersburg State University (Project ID No. 94031444).

%


\end{document}